\documentclass[a4paper]{article}

\usepackage{INTERSPEECH2020}
\usepackage{amsmath,graphicx,amssymb}
\usepackage{subcaption}
\usepackage[ruled,linesnumbered]{algorithm2e}
\usepackage{soul}

\usepackage{multirow, url}

\usepackage[activate]{microtype}
\newcommand*{\SL}[1]{\textcolor{black}{#1}}
\newcommand*{\TR}[1]{\textcolor{black}{#1}}

\usepackage[dvipsnames]{xcolor}

\DeclareMathOperator{\EX}{\mathbb{E}}

\title{Deep Reinforcement Learning with Pre-training for Time-efficient Training of Automatic Speech Recognition}

\name{Thejan Rajapakshe$^{1}$, Siddique Latif$^{1,2}$, Rajib Rana$^{1}$, Sara Khalifa$^{2}$, Bj\"{o}rn W.\ Schuller$^{3,4}$}

\address{ $^1$University of Southern Queensland, Australia\\
  $^2$Distributed Sensing Systems Group, Data61, CSIRO Australia\\
  $^3$GLAM -- Group on Language, Audio \& Music, Imperial College London, UK\\
  $^4$ZD.B Chair of Embedded Intelligence for Health Care \& Wellbeing, University of Augsburg, Germany}

\email{Thejan.Rajapakshe@usq.edu.au}
\begin{document}
%
\maketitle
\begin{abstract}
Deep reinforcement learning (deep RL) is a combination of deep learning with reinforcement learning principles to create efficient methods that can learn by interacting with its environment. This has led to breakthroughs in many complex tasks, such as playing the game ``Go'', that were previously difficult to solve. However, deep RL requires significant training time making it difficult to use in various real-life applications such as Human-Computer Interaction (HCI). In this paper, we study 
pre-training in deep RL to reduce the training time and improve the performance of Speech Recognition, a popular application of HCI. 
To evaluate the performance improvement in training we use the publicly available ``Speech Command'' dataset, which contains utterances of 30 command keywords spoken by 2,618 speakers. Results show that pre-training with deep RL offers faster convergence compared to non-pre-trained RL while achieving improved speech recognition accuracy.

\end{abstract}

\section{Introduction}
Reinforcement
Learning (RL) follows the principle of behaviourist  psychology and learns 
similarly
as a child learns to perform a new task. RL has been 
repeatedly successful in the past~\cite{singh2002optimizing,tesauro1995temporal}, however, the successes were mostly limited to low-dimensional problems. In recent years, deep learning has significantly advanced the field of RL, with the use of deep learning algorithms within RL giving rise to the field of ``deep reinforcement learning''. Deep learning enables RL to operate in high-dimensional state and action spaces and 
can now be used for complex decision-making problems~\cite{arulkumaran2017brief}.

Deep RL algorithms have been applied to video or image processing domains spanning video games~\cite{silver2016mastering,mnih2015human} to indoor navigation~\cite{zhu2017target}. Very few studies have explored the promising aspects of deep RL in the field of audio processing particularly, in speech processing \cite{latif2020deep}. 
In this paper, we focus on this under-researched topic. Specifically, we conduct a case study of the feasibility of deep RL for 
automatic 
speech command classification. 

A major challenge of deep RL is that it often requires a prohibitively large amount of training time and data to reach a reasonable accuracy, making it inapplicable in real-world settings~\cite{cruz2017pre}. Leveraging humans to provide demonstrations (known as learning from demonstration (LfD)) in RL has recently gained traction as a possible way of speeding up deep RL~\cite{vinyals2017starcraft,hester2018deep,kurin2017atari}. In  LfD, 
actions demonstrated by the human are considered as the ground truth labels for a given input game/image frame. 
An agent closely simulates the demonstrator's policy at the start, and later on, 
learns to surpass the demonstrator~\cite{cruz2017pre}. However, LfD holds a distinct challenge, in the sense that it often requires the agent to acquire skills from only a few demonstrations and interactions due to the time and expense of acquiring them~\cite{calinon2018learning}. Therefore,
LfDs are generally not scalable, especially for high-dimensional problems.

\SL{Pre-training the underlying deep neural network is another approach to speed up training in deep RL. It enables the RL agent to learn better features which leads to better performance without changing the policy learning strategies \cite{cruz2017pre}.} 
In supervised methods, pre-training helps regularisation and enables faster convergence compared to randomly initialised networks \cite{yu2010roles}. Various studies (e.\,g., \cite{thomas2013deep,liu2014graph}) have explored pre-training in speech recognition and achieved improved results. However, pre-training in deep RL is hardly explored in the area of speech recognition. In this paper, we propose a deep RL framework for speech recognition and evaluate the performance of  pre-training to reduce the training time.



\section{Related Work}
\SL{Deep RL often requires prohibitively large amounts of training time and data to achieve a reasonable performance, which makes it unsuitable for real-world applications. Pre-training  in deep RL is useful to speedup the training process and to reduce the requirement of a large amount of data \cite{zhang2018pretraining}.
Authors in \cite{blau2018improving} use sparse variational dropout regularisation for pre-training RL and show that pre-training allows an RL algorithm to learn optimal policies for high-dimensional continuous control problems in a practical time frame. In~\cite{abtahi2011deep}, the authors combine Deep Belief Networks (DBNs) with RL to take advantage of the unsupervised pre-training phase in DBNs and then use the DBN as the opening point for a neural network function approximator. The authors in~\cite{anderson2015faster} demonstrate that a pre-trained hidden layer architecture can reduce the time required to solve RL problems.} \SL{While these studies show the promise of using pre-trained deep RL, they are not in audio domains. 
The feasibility of pre-training RL for the audio is not yet well understood. In this paper, we investigate the usability of pre-training of deep RL for speech recognition.}

\SL{Although we could not find studies using pre-training of RL for for audio, some studies used pre-training in speech research for Deep Learning (DL) models. Thomas et al.\  \cite{thomas2013deep} utilised pre-training for Deep Neural Networks (DNN), where they achieved excellent results for speech recognition, by utilising only 1 hour of transcribed training data. Some studies (e.\,g., \cite{imseng2014exploiting}) also achieved promising results for cross-lingual acoustic data using pre-training in deep learning neural networks. In contrast to these studies, we use pre-training for speech-based systems in deep RL setting.}



\section{Methodology}
\label{sec:methods}
\TR{Feature Learning and Policy Learning are the main two sub-tasks of Deep RL~\cite{cruz_jr_pre-training_2019}.}
\SL{To investigate pre-training in deep RL setting, we propose a model for speech command recognition whose details are explained below. 
}

\subsection{Pre-Training}
\TR{Understanding the impact of pre-training on the performance of RL is the primary aim of this study.
Using the Speech Command dataset, we trained a conventional supervised DNN model and the model parameters were used to initialise the policy network (see Section~\ref{sec:pn}) of the Deep RL. We refer to this process as pre-training. Pre-training helps the model to converge quickly and help improve the accuracy of inference for unseen data during the RL execution.}

\subsection{Deep Reinforcement Learning Framework}
\begin{figure}[t!]
    \centering
    \includegraphics[width=0.75\linewidth]{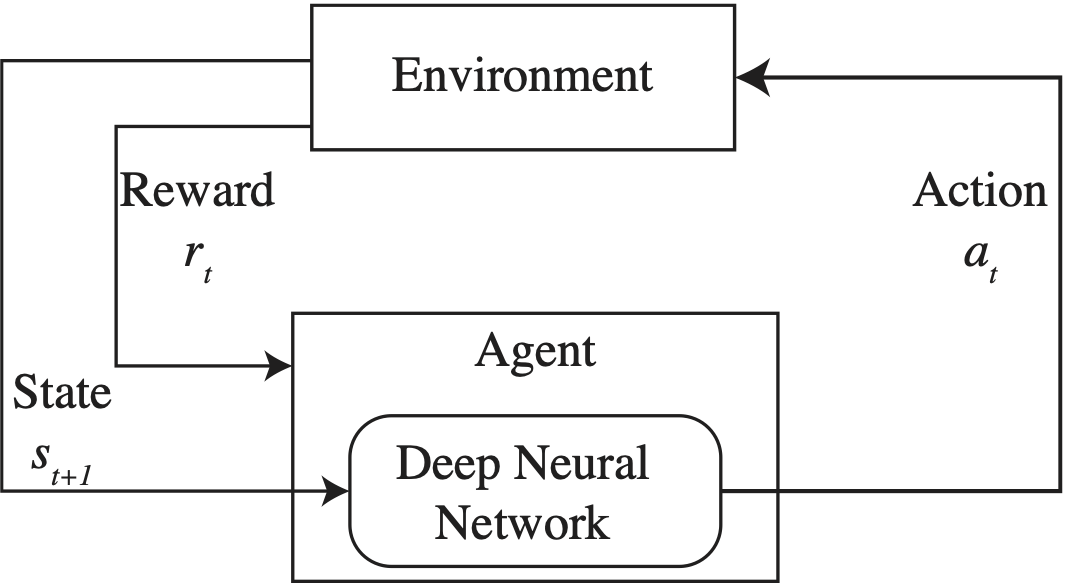}
    \caption{Framework of the proposed Deep RL. }
    \label{fig:rl_framework}
\vspace{-5mm}
\end{figure}
The reinforcement learning framework mainly consists of two major entities namely ``agent'' and ``environment''. The \textit{action} decided by the agent is executed on the environment and it notifies the agent with the reward and next state in the environment. In this work, we focus on deep RL that involves a DNN structure in the agent module to resolve the action taken by observing the state which is illustrated in Figure \ref{fig:rl_framework}. We modelled this problem as a Markov decision process (MDP) \cite{tetreault2008reinforcement}. This can be considered as a tuple $(S,A,P,R)$, where $S$ is the state space, $A$ is the action space, $P$ is the state transition policy, and $R$ is the reward function. Since the core goal of this problem is classification, we modelled the MDP in such a way that the predicted classes are to be as actions, $A$, and the states, $S$ are the features of each audio segment in a batch of size $\eta$. An action decision is carried out by an RL agent which receives a reward ($r_{t}$) using the following reward function: 
\begin{equation}
    r_{t}=
    \begin{cases}
      +1, & \text{if}\ a_{t} = g_{t} \\
      -1, & \text{otherwise}, 
    \end{cases}
\end{equation}

where $g_{t}$ is the ground truth value of the specific speech utterance. We modelled the probability of actions using the following equation: 
\TR{
    \begin{equation}
        a_{t} = \text{argmax}(\text{softmax}(W^{a}\cdot S_{t} +b^{a})), 
    \end{equation}}
\TR{
where $a_{t}$ is the class index of the maximum probability, $g_{t}$ is the ground truth value of the specific speech utterance}
and $W^{a}$ and $b^{a}$ are the weight and bias values. $S_{t}$ is the output from the previously hidden layer.


\SL{The target of the RL agent is to maximise the expected return using the following policy:}
\begin{equation}
\label{obj}
    J_{a} (\theta_a,\theta_s) = \EX_{\pi(a_{t}|s_{t}; \theta_a,\theta_s)}[r_{t}], 
\end{equation}
where $\pi(a_{t}|s_{t}; \theta_a,\theta_s)$ is the policy of agent, and $r_{t}$ is the expected reward return at state $t$. \SL{To update the policy, we utilise the policy network. Details on the policy network are presented next.  }

\subsection{Policy Network}
\label{sec:pn}
\begin{figure}[t!]
    \centering
    \includegraphics[width=\linewidth]{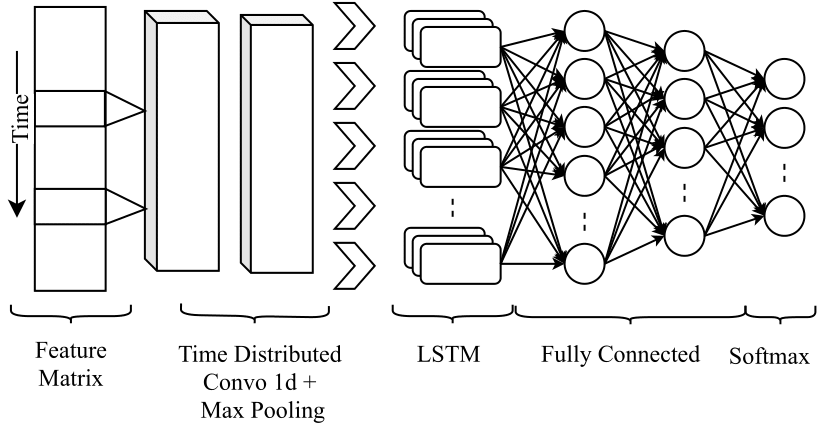}
    \caption{Speech command recognition model architecture.}
    \label{fig:speech_command_model}
\end{figure}

The policy network model consists of a speech command recognition model as shown in  figure~\ref{fig:speech_command_model}.
\SL{The policy network learns to generate a definite output for a particular input in an RL algorithm. In this work, the policy network takes speech features as input state and recognises the spoken command. For this, we use a deep network consisting of convolutional (CNN) and Long Short Term Memory (LSTM) layers. Our choice of CNN-LSTM is motivated by their ability to learn both temporal and frequency components of speech signals \cite{latif2020deep}. 
An LSTM cell in recurrent neural networks (RNNs) is a memory unit for learning the temporal structure of sequential data \cite{latif2018phonocardiographic}, and CNNs are strong in 
diminishing frequency variations \cite{latif2020multi}. We assemble CNN layers on top of an LSTM RNN layer.} The outputs from the LSTM RNN layer are then passed on to fully connected layers to learn discriminative features during training \cite{latif2019direct}. In this way, our proposed policy network is empowered by convolutional layers for learning high-level abstraction, an LSTM RNN layer to capture long-term temporal context, and finally fully connected layers for learning discriminative representation. 



\subsubsection{Trainable Model}

\begin{figure}[t!]
    \centering
    \includegraphics[width=0.75\linewidth]{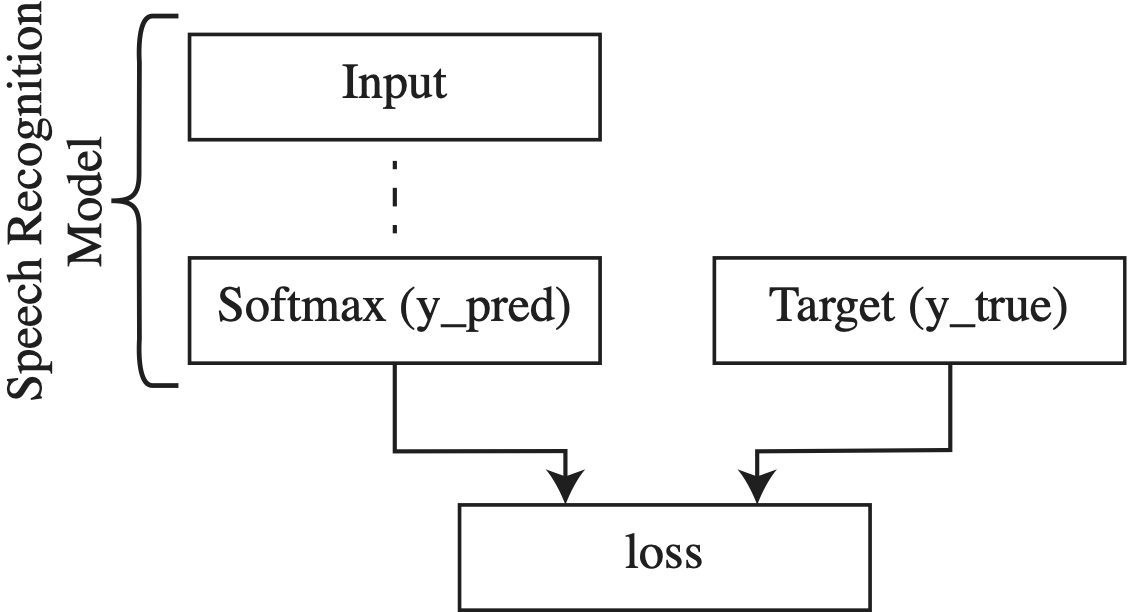}
    \caption{Trainable model architecture.}
    \label{fig:trainabl_model}
\vspace{-4mm}
\end{figure}

To calculate the accuracy, we created a separate network by stacking the loss function on top of the output of the policy network and the ``target model'' as shown in Figure~\ref{fig:trainabl_model}. The target model is of the same architecture of the policy network and it updates weights from the policy network once every \TR{500 episodes}. \TR{This target model is used to infer the target values ($Y\textunderscore true$).} 

\subsubsection{REINFORCE Algorithm}
The REINFORCE algorithm is used to approximate the gradient to maximise the objective function $J(\theta_{a},\theta_{s})$ mentioned in Equation \ref{obj}. 

\begin{algorithm}[t!]
\SetAlgoLined
 initialise state space\;
 initialise policy network model\;
 pre-train policy network\; 
 retrieve initial state $s_{1}$\;
 \For{$i \leftarrow 1$ \KwTo $N_{E}$}{
    initialise $E_{s}, E_{a}, E_{r}$\;
    \While{!d}{
        $a_{i} \leftarrow $ get action($s_{i}$)\;
        $s_{i+1}$, $r_{i}$, $d \leftarrow execute(a_{i}$)\;
        $E_{s} \leftarrow s_{i} + E_{s}$\;
        $E_{r} \leftarrow r_{i} + E_{r}$\;
        $E_{a} \leftarrow a_{i} + E_{a}$\;
    }
    $train(E_{s}, E_{a}, E_{r})$\;
 }
 \caption{REINFORCE algorithm implementation}
 \label{alg:rl_algo}
\end{algorithm}

\begin{algorithm}[t!]
\KwIn{History\textless Rewards,States,Actions\textgreater}
\SetAlgoLined
$Y\textunderscore true \leftarrow $ predict from the target model (States)\;
$Y\textunderscore pred \leftarrow $ policy network output\;
$R \leftarrow$ Reward + Discounted Reward\;
$loss \leftarrow $ lambda:  clipped error($Y\textunderscore pred, Y\textunderscore true$)\;
Gradient descent on $trainable\textunderscore model$ inputs=$[States,Y\textunderscore true]$, output=$[R]$ with loss function $loss$\;
\caption{Training the policy network}
\label{alg:train_netowrk}
\end{algorithm}

Algorithm~\ref{alg:rl_algo} describes the algorithmic steps followed throughout the RL action prediction process, where $N_{E}$ indicates the maximum number of episodes to run (10,000 experiments).
At the beginning of each episode, a subset of the initial dataset (N=50) is selected randomly as the state space $S$.
$S_{i}$ is the state at instant $i$, $a_{i}$ is the predicted action for the $S_{i}$ at the $i^{th}$ instant, $r_{i}$ is the reward obtained by executing the predicted action $a_{i}$, $d$ is a boolean flag indicating the end of an episode, where the end of the episode is decided when $i$ reaches the step size $\eta$ (50). $E_{s}, E_{a}, E_{r}$ are arrays collecting the values of $s_{i}$, $a_{i}$, $r_{i}$ for each step, which is consumed by the policy model's training method $train$ described in Algorithm~\ref{alg:train_netowrk}. Training is carried out at the end of each episode and ``target model'' update its weights from policy network after every 200 episodes 




\section{Experimental Setup}
\subsection{Dataset}


\SL{To evaluate the proposed framework, we used the publicly available Speech Commands Dataset. The speech commands dataset \cite{warden_speech_2018} contains utterances of 30 command keywords spoken by 2,618 speakers. Each utterance represents a one-second file with a sampling rate of 16\,kHz. This dataset contains mainly two subsets of command keywords, namely ``main commands'', and ``sub commands''. Table~\ref{tab:sc_dataset} shows the distribution of the 30 keywords among the two subsets. }
\begin{table}[h]
    \centering
    \caption{Distribution of keywords in the Speech Commands Dataset}
    \begin{tabular}{p{8em}p{13em}}
        \hline
        Subset & Commands \\
        \hline
         Main Commands & one, two, three, four, five, six, seven, eight, nine, down, go, left, no, off, on, right, stop, up, yes, zero  \\\hline
         Sub Commands & bed, bird, cat, dog, happy, house, Marvin, Sheila, tree, wow
         \\\hline
    \end{tabular}
    
    \label{tab:sc_dataset}
\end{table}
\TR{Only 10\% of the speech commands dataset was separated for the pre-training step and the remaining 90\,\% was used by the RL environment.}

\subsection{Feature Extraction}
We use Mel Frequency Cepstral Coefficients (MFCC) to represent the speech signal. MFCCs are very popular features and widely used in speech and audio analysis \cite{latif2019direct,davis1980comparison}. We extract 40 MFCCs from the Mel-spectrograms with a frame length of 2,048 and a hop length of 512 using Librosa \cite{mcfee2015librosa}.

\begin{figure*}[!ht]
    \centering
    \includegraphics[trim=0.2cm 0.2cm 0.2cm 0.2cm,clip=true,width=\textwidth]{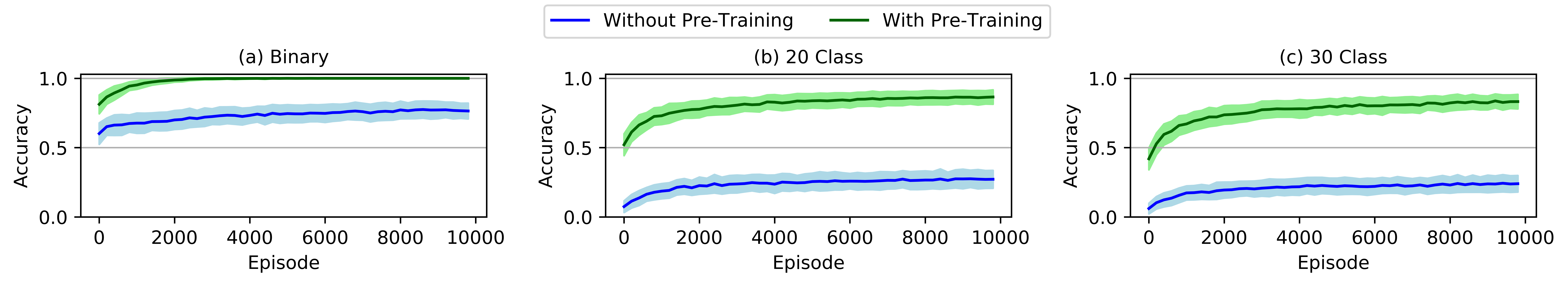}
    \caption{Performance evaluations of the model on three different scenarios}
    \label{fig:performence}
\end{figure*}
\begin{figure*}[!ht]
    \centering
    \includegraphics[trim=0.2cm 0.2cm 0.2cm 0.2cm,clip=true,width=\textwidth]{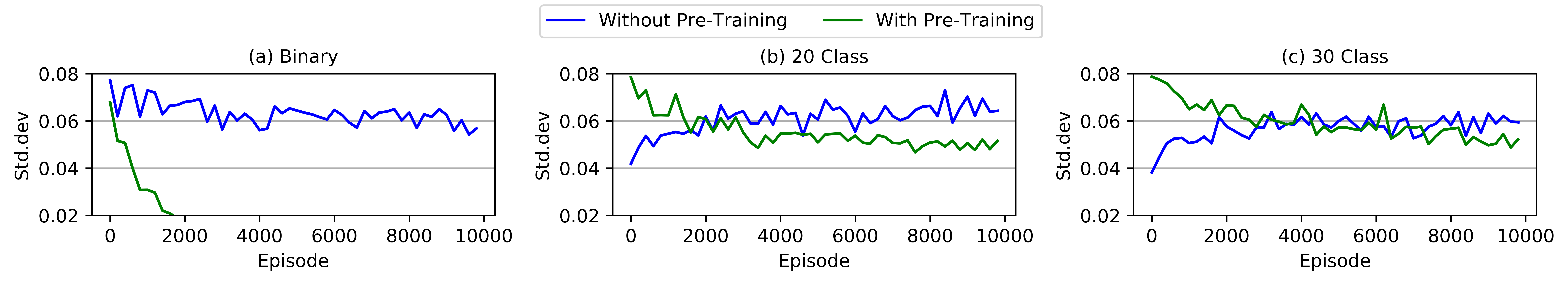}
    \caption{Standard Deviation of the accuracy with the episode on three different scenarios}
    \label{fig:std_dev_episode}
\end{figure*}

\subsection{Model Recipe}
We use the Tensorflow library to implement the policy network, which is a combination of CNN and LSTM. The initial layers are 1d convolution layers wrapped in time distributed wrappers with filter sizes of 16 and 8,  respectively, followed by a max-pooling layer. The feature maps are then passed to an LSTM layer of 50 cells for learning the temporal features. A dropout layer of dropout rate 0.3 is used for regularisation. Finally, three fully connected layers of 512, 256, and 64 units respectively are added before the softmax layer. 

The input to the model is a matrix of $n \times f$, where $n$ is the number of MFCCs (40), and $f$ is the number of frames (87) in the MFCC spectrum. 
We use a stochastic gradient descent optimiser 
with a learning rate of $10^{-4}$.
The \TR{pre-training steps were carried out with stochastic gradient descent as the optimiser with a learning rate of 0.001. The model was trained for 10 epochs with a batch size of 8 and 10\,\% as validation split. }

\TR{The ``target model'' does not update the weights during the training phase but updates the weights after every 200 episodes with the weights from the policy network. The ```loss'' tensor in the trainable model takes outputs from the ``target model'' and policy network as inputs, then calculates the loss at the end of each episode. This loss is minimised through the Adam optimiser. This adjusts the weights of the policy network towards the optimum.}

Accuracy of each episode $i$ ($H_i$) is calculated by Equation~\ref{eq:accuracy}. Where $\eta_\text{correct}$ is the number of correct predictions and $\eta$ is the total number of steps per episode. We use $\eta= 50$.
\begin{equation}
    H_{i} = \frac{\eta_{\text{correct}}}{\eta}
    \label{eq:accuracy}
\end{equation}



\section{Results}
To benchmark the results of the RL accuracy, we train a DNN with the same model configuration as of our policy network. We use 80\% of data for training and 20\% for testing. We use Stochastic gradient descent as the optimizer, where we use learning rate $10^{-4}$ and batch size 32. We present the comparison results in Table~\ref{tab:benchmark_accuracy} .

\begin{table}[t!]
\centering
\caption{Benchmarking results using the model in supervised training. Given are the different number of classes.}
\begin{tabular}{l|r|r|r}
\hline
Classes       & Binary                                                                              & 20 Classes                                          & 30 Classes                                            \\ \hline
Accuracy (\%) & \begin{tabular}[c]{@{}c@{}}87.35\end{tabular}  & \begin{tabular}[c]{@{}c@{}}81.03\end{tabular} & \begin{tabular}[c]{@{}c@{}}78.88\end{tabular} \\ \hline
\end{tabular}
\label{tab:benchmark_accuracy}
\end{table}

Experiments were carried out to identify the impact of pre-training on the training-time and accuracy of the RL Agent. Three subsets of speech command datasets were selected, namely ``binary'', ``20 class'', and ``30 class''.  The binary subset contains only the speech commands ``left'' and ``right''. 20 classes and 30 classes subsets contain "main" commands and the merge of ``main'' and ``sub'' commands, respectively in the ``Speech Command'' dataset.


We perform experiments using the proposed deep RL model on each subset and report the results in Tables \ref{tab:initial_accuracy_improvement}. 
\TR{Table~\ref{tab:initial_accuracy_improvement} provides the mean accuracy of 200 initial episodes for ``with'' ($w/$) and ``without'' ($w/o$) pre-training. We observe that for all  classification subsets, non-pre-trained RL gain considerably lower accuracy for the initial 200 episodes. However, while using pre-training, using the same number of episodes we achieve significantly higher accuracy. This essentially shows that using pre-training we are able to reduce the training time significantly.}

Table~\ref{tab:initial_accuracy_improvement} also shows the mean accuracy of the latest 5 episodes after 10,000 episodes for the ``with'' ($w/$) and ``without'' ($w/o$) pre-training scenarios. \TR{Pre-trained RL after 10,000 episodes suppresses the benchmark results on every experiment reported in Table~\ref{tab:benchmark_accuracy}. The improvement column ```$\Delta$'' shows the increment of the accuracy of the ``with pre-training'' with respect to the ``without pre-training'' scenarios. Each improvement is significant, which further strengthen our findings that pre-train can reduce the training time for Deep RL.}




\begin{table}[t]
\scriptsize
\centering
\caption{Improvement of the accuracy in \% with (w/) and without (w/o) pre-training at 200 initial episodes and after 10,000 episodes. Also shown is the difference $\Delta$.}
\resizebox{\linewidth}{!}{%
\begin{tabular}{r|rrr|rrr}
\hline
\multirow{2}{*}{\# Classes} & \multicolumn{3}{c}{Initial 200 episodes} & \multicolumn{3}{|c}{After 10000 episodes} \\
 & $w/o$ & $w/$ & $\Delta$ & $w/o$ & $w/$ & $\Delta$ \\ \hline
2 & 60.13 & 81.24 & 21.11 & 80 & \textbf{100} & 20 \\
20 & 7.43 & 52.11 & 44.68 & 25.71 & \textbf{87.76} & 62.04 \\
30 & 6.05 & 41.92 & 35.87 & 26.12 & \textbf{79.59} & 53.47 \\ \hline
\end{tabular}%
}
\label{tab:initial_accuracy_improvement}
\vspace{-6mm}
\end{table}

\TR{To further demonstrate the improvement in training time, the accuracy of the episodes was plotted against the episode number and presented in Figure~\ref{fig:performence}.
One can observe that the pre-training has increased the overall accuracy in each of the 3 experiments. Also, when the rate of change of accuracy is observed within the initial 2000 episodes it can be seen that the rate of change of accuracy is increased in all the pre-trained experiments. This infers that the number of episodes needed to achieve a defined accuracy is reduced by pre-training. Hence the efficiency is improved.}

Lower standard deviation indicates higher consistency. Standard deviation of the accuracy is plotted against the episode in Figure~\ref{fig:std_dev_episode} and it can be observed that the standard deviation has decreased rapidly in all the pre-trained experiments. This observation deduces that the pre-training improves the consistency of the predictions earlier.




\section{Conclusions}
In this paper, we propose the use of pre-training in deep reinforcement learning for speech recognition. The newly introduced framework uses pre-training for feature learning in a reinforcement learning problem. The learned feature knowledge through pre-training is used by Policy Learning during the reinforcement execution to achieve higher accuracy within a reduced time.
We evaluate the proposed RL model using the Speech Command dataset for three different classification scenarios, which include binary (two different speech commands), and 20 and 30 class tasks. The results show that pre-training improves the time-efficiency of RL, helping to achieve considerably better results in a significantly smaller number of episodes compared to without using pre-training for RL. 




\end{document}